\documentclass[10pt,letterpaper,twocolumn]{article} 

\usepackage{ol2}
\usepackage[draft]{hyperref}
\usepackage{amsmath}

\begin{document}

\twocolumn[ 

\title{Frequency down-conversion for a quantum network}


\author{Yu Ding and Z. Y. Ou$^*$}

\address{Department of Physics, Indiana University-Purdue University
Indianapolis, \\402 N. Blackford Street, Indianapolis, IN 46202,
USA \\
$^*$Corresponding author: zou@iupui.edu}

\begin{abstract}
By using parametric down-conversion process with a strong signal field injection, we demonstrate coherent frequency down-conversion from a pump photon to an idler photon. Contrary to a common misunderstanding, we show that the process can be free of quantum noise. With an interference experiment, we demonstrate that the coherence is preserved in the conversion process. This may lead to a high fidelity quantum state transfer from high frequency photon to low frequency photon and connects a missing link in a quantum network. With this scheme of coherent frequency down-conversion of photons, we propose a method of single-photon wavelength division multiplexing.
\end{abstract}


]

\noindent Photons are generally believed to be good quantum information
carriers for transmission and are dubbed the term ``flying qubits"
whereas atoms are best for storing and processing the quantum
information. Therefore, a quantum network usually consists of
nodes made of atoms and connected by photons \cite{cir}. In the
network, quantum information is constantly transferred between
photons and atoms and transmitted between atoms via photons. Because quantum information is
sensitive to losses, minimum losses are required in the network. However, in current technology, atoms interact best with photons of wavelength around $0.8 \mu m$ \cite{cir,kim} whereas optical communication system has low losses at $1.6 \mu m$ \cite{kao}. So there is a mismatch between the atomic transition wavelength and the optical transmission wavelength. Thus, it is necessary to convert photons from one wavelength to another in order to set up the quantum network.

Quantum information transfer between atoms and photons has been
realized \cite{kim2} in a near-resonance Raman system  based on the electromagnetically induced
transparency (EIT) effect \cite{fle,hau2,wal1,wal2,kim4,mir}.
For the photonic state transfer between different wavelengths,
frequency up-conversion process was realized many years ago
\cite{kum,de,van,al,ro,gisin}, based on sum-frequency generation but none was reported for frequency
down-conversion. This is so because current research focus is on
up-converting photons at optical communication wavelength ($\sim 1.6
\mu m$) to shorter wavelength ($\sim 0.8 \mu m$) for photon counting in quantum cryptography \cite{gisin} -- detectors at $1.6
\mu m$ are just far more noisy than those at  $0.8 \mu m$. Moreover, the noise behavior in parametric dow-conversion process \cite{cave} leads to the belief that it is impossible to have high fidelity quantum state transfer in frequency down-conversion process \cite{de}.  So, frequency down-conversion seems to be the missing link in a quantum network. Frequency conversion was also achieved in four-wave mixing via Bragg scattering in fiber\cite{mc,rad}. But because phonons are involved, the frequency shift is relatively small.

In this letter, we study a parametric down-conversion scheme for photon frequency down-conversion and quantum information
transfer. We will demonstrate that contrary to some misunderstanding about parametric down-conversion, noise-free photon frequency down-conversion is achievable in this process. We implement a proof-of-principle experiment and show that quantum coherence is preserved in the process.

The parametric frequency down-conversion process is usually described by the Hamiltonian \cite{lou,lou2,shen,boyd}:
\begin{eqnarray}
\hat H_{PA}=i\hbar \eta A_p \hat a_s^{\dag}\hat a_i^{\dag}
-i\hbar \eta^* \hat a_s \hat a_iA_p^{*},\label{HPA}
\end{eqnarray}
where $A_p$ is the strong pump field and usually is treated as a classical field.  ``$s,i$" stand for ``signal" and ``idler" fields for historic
reason. This leads to the evolution equation for parametric amplifier:
\begin{eqnarray}
\hat a_s^{(out)} = G \hat a_s
+ g\hat a_i^{\dag},~~~~~\hat
a_i^{(out)}=  G \hat a_i  + g\hat
a_s^{\dag},\label{Ba}
\end{eqnarray}
where $G\equiv \cosh |\eta A_p|\tau$ is the amplitude gain and $g\equiv -e^{j\varphi_p}\sinh |\eta A_p| \tau $.
The appearance of the $\hat a^{\dag}$-terms in
Eq.(\ref{Ba}) leads to spontaneous quantum noise and the belief that frequency down-conversion is noisy and cannot preserve quantum coherence\cite{cave,de}.

On the other hand, there is another regime of operation in which we inject strong signal field. This regime is often ignored because of gain saturation: the amplification of the strong signal field requires more energy from the pump field which eventually will be depleted. In this regime of operation, the spontaneous emission for frequency down-conversion is
negligible and we may achieve a noiseless frequency down-conversion
of photons. We will show this more rigorously in the following.

When the pump field is depleted, we can no longer use the evolution equations in Eq.(\ref{Ba}). We need to start with a Hamiltonian for three-wave mixing:
\begin{eqnarray}
\hat H_{3W}=i\hbar \eta \hat a_p \hat a_s^{\dag}\hat a_i^{\dag}
-i\hbar \eta^* \hat a_s \hat a_i\hat a_p^{\dag}.\label{HIII}
\end{eqnarray}
Here the pump field is also treated quantum mechanically.
If the ``signal" field is very strong, as in our case here, we can treat it as a classical field and replace it with a constant $A_s$.
The pump field and the ``idler" field are quantum fields. So the Hamiltonian in
Eq.(\ref{HIII}) becomes
\begin{eqnarray}
\hat H_{FC}=i\hbar \eta \hat a_p \hat a_i^{\dag}A_s^* -i\hbar \eta^*
A_s \hat a_i\hat a_p^{\dag}.\label{HC}
\end{eqnarray}
The evolution of the pump and the ``idler" fields is exactly that for a beam splitter \cite{tei} and is given by \cite{lou,lou2}
\begin{eqnarray}
&\hat a_p^{(out)} = \hat a_p\cos|\eta A_s| \tau +
e^{j\varphi_s}\hat a_i\sin |\eta A_s|\tau ,\cr & \hat a_i^{(out)}=
\hat a_i\cos |\eta A_s| \tau- e^{-j\varphi_s} \hat a_p\sin |\eta
A_s|\tau ,\label{Bf}
\end{eqnarray}
where $e^{j\varphi_s}= \eta^* A_s/|\eta A_s|$ and $\tau$ is the interaction time.  When $|\eta A_s|
\tau = \pi/2$, we have $\hat a_p^{(out)} =  e^{j\varphi_s}\hat
a_i$ and $\hat a_i^{(out)} =  - e^{-j\varphi_s}\hat a_p$. So we
may achieve quantum field conversion between $\hat a_i$ and $\hat
a_p$ with unit conversion efficiency. Note that both  the frequency down-conversion of
$\hat a_p \rightarrow \hat a_i^{(out)}$ and up-conversion of $\hat a_i \rightarrow \hat a_p^{(out)}$  co-exist and it all depends on what the input field is.

In the non-ideal case when $|\eta A_s|
\tau < \pi/2$, Eq.(\ref{Bf}) is simply a beam splitter equation, which mixes the two inputs together coherently. The transmissivity is simply $t \equiv \cos |\eta A_s|
\tau$ and the reflectance $r\equiv \sin |\eta A_s|
\tau$.

\begin{figure}
\centerline{\includegraphics[width = 3.2in]{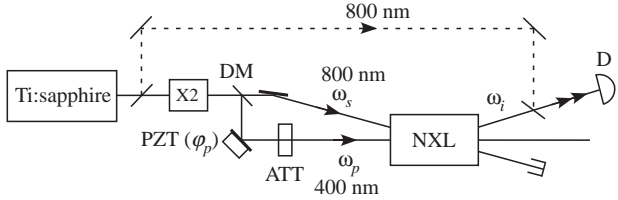}}
\caption{\em Schematics for the demonstration of frequency down-conversion. {\rm X2}: frequency doubling; DM: dichroic mirror; PZT: piezoelectric transducer; ATT: attenuator; D: photo-detector.}
\end{figure}

Next, we describe a proof-of-principle demonstration experiment for photon frequency down-conversion. The experimental sketch is shown in Fig.1. The source of light in our experiment is a femto-second Ti:sapphire laser with a modest power (300mW) and a repetition rate of 80 MHz. The ``signal" field is directly from the laser with a wavelength of 850 nm. The pump field is from the attenuation of the frequency doubling of the laser. When the two pulses are overlapping at a nonlinear crystal (LiIO$_3$) in a non-collinear fashion with a proper angle for phase matching (Fig.1), a field with frequency difference ($\omega_i = \omega_p-\omega_s$) is generated at another direction. In the first experiment, we check the linearity of the detected ``idler" field as a function of the attenuation factor on the incoming pump field, as predicted from Eq.(\ref{Bf}). The result is shown in Fig.2 in logarithmic scale. It can be seen that in a range of 2 orders of magnitude, the detected signal follows well the linear dependence. The last point at the low end is due to the limit of the detector sensitivity. By a direct measurement of input power at the pump and output power of the idler, we estimate the photon conversion efficiency is about 1\%. The low efficiency is a result of an inefficient nonlinear crystal that we pulled out of shelf for the proof-of-principle experiment. More efficient nonlinear materials like PPLN can significantly improve the conversion efficiency.

\begin{figure}
\centerline{\includegraphics[width = 3in]{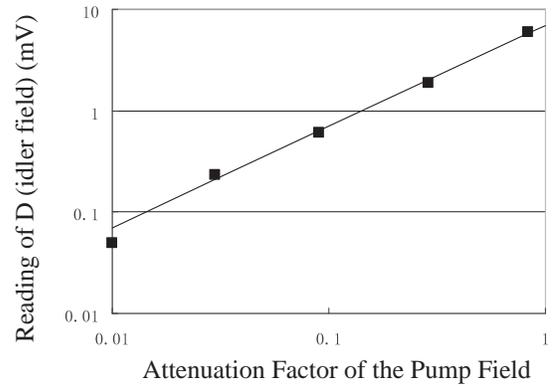}}
\caption{\em The detected signal size of the ``idler" field as a function of the attenuation factor of the pump field or the transmission coefficient of the attenuator ATT. The solid line is $y=ax$ with $a=7$ for best fit.}
\end{figure}

\begin{figure}
\centerline{\includegraphics[width = 3in]{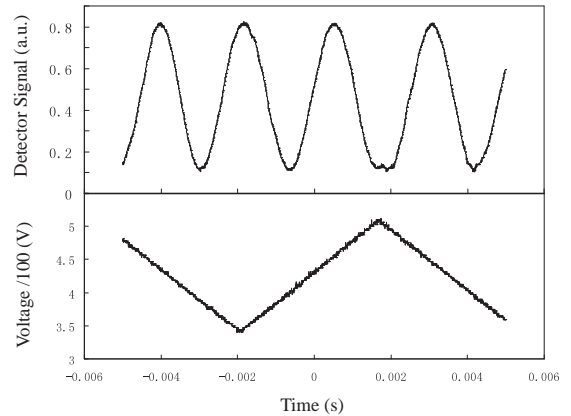}}
\caption{\em Upper trace: interference fringe between the generated ``idler" field and the original laser field. Lower trace: the ramp signal (/100) from the high voltage applied to the PZT for phase scan of the pump field.}
\end{figure}

According to Eq.(\ref{Bf}), the output-input relation is that for a beam splitter, which preserves the phase coherence. So, we check next the phase preservation in the conversion from the pump field to the ``idler" field. Here, we use the original laser as a reference and beat the generated ``idler" field with a small portion split from the laser (dashed line in Fig.1). Since the down-conversion process that we are using is a degenerate one with $\omega_s= \omega_i= \omega_p/2$, the generated ``idler" field has the same frequency as the laser. So, we should be able to observe the interference effect. Fig.3 shows a trace of the detected combined field at detector D. The phase scan is on the pump field and is achieved by applying a ramp voltage on a piezoelectric transducer (PZT in Fig.1). The sinusoidal change in the detected signal shows the interference effect between the generated ``idler" field and the laser field but with a phase dependence from the pump field. This clearly demonstrates the coherent photon conversion from the pump field to the ``idler" field.

The frequency down-conversion scheme can be used to create a superposition of multiple frequency components from an input of single photon of one frequency component, equivalent to
the wavelength division multiplexing (WDM) technique
widely used in classical optical fiber communication.
Such a technique combines multiple-wavelength components from different lasers into one field. The source so created has different channels that are independent of each other. So different information  can be simultaneously transmitted over a single fiber, thus increasing
the channel capacity of the fiber.  In quantum communication, we may apply the same multiplexing technique on quantum information carriers, namely, photons. But since quantum information requires preservation of quantum superposition, different channels may not be independent to each other.
Furthermore, in quantum key distribution (QKD) \cite{gisin05}, a single-photon source is preferred to defeat the photon splitting attack. So, for multi-channel QKD, we need a single-photon source with WDM. In the WDM-QKD system, we can use multiple single-photon sources with different frequencies, just like in classical optical communication. However, single-photon sources are often from a single quantum system such as atom or ion, which gives rise to a single frequency.  And sometimes we may also have only one such kind of system due to limited resources.

\begin{figure}
\centerline{\includegraphics[width = 3.0in]{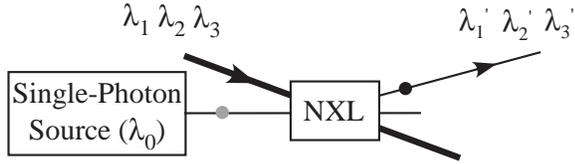}}
\caption{\em Single-photon wavelength division multiplexing.}
\end{figure}

Using the frequency down-conversion scheme discussed earlier, we can create multiple frequency components from only one single-frequency source and implement the wavelength division multiplexing technique on single photons. The idea is to use light source of many wavelengths as the ``signal" field to drive the three-wave mixing process (Fig.4).   Due to energy conservation: $\omega_p
= \omega_s+\omega_i$, the down-converted ``idler" field will have different frequencies $\omega_{i1},
\omega_{i2}, \omega_{i3}, ...$ corresponding to the different frequencies of the ``signal" field. The system will behave like a multiple of beam splitters for the input pump field. If  the input pump field is in a single-photon state $|1\rangle_p$,
the output will be a superposition state of many frequencies:
$
|1\rangle_p \rightarrow \sum_k c_k|1_{\omega_k}\rangle, $
realizing single-photon wavelength division multiplexing. The strong ``signal" field of multiple wavelengths can be from a mode-locked laser \cite{ye} where all frequency components are coherent to each other so that the photon so-produced is in a superposition state of different frequencies.  Such a wavelength division multiplexing scheme for quantum source preserves quantum entanglement.

\end{document}